# Unusual Effects of Be doping in the Iron Based Superconductor FeSe

J. S. Kim,[1] D. VanGennep,[1] J. J. Hamlin,[1] X. Wang,[2] A. S. Sefat,[2] and G. R. Stewart[1]

[1]Dept. of Physics, University of Florida, Gainesville, FL 32611

[2]Oak Ridge National Laboratory, Oak Ridge, TN 37831

Email: stewart@phys.ufl.edu

**Abstract**

Recent superconducting transition temperatures ($T_c$) over 100 K for monolayer FeSe on $SrTiO_3$ have renewed interest in the bulk parent compound. In $KCl:AlCl_3$ flux-transport-grown crystals of $FeSe_{0.94}Be_{0.06}$, $FeSe_{0.97}Be_{0.03}$ and, for comparison, FeSe, this work reports doping of FeSe using Be – among the smallest of possible dopants, corresponding to an effective 'chemical pressure'. According to lattice parameter measurements, 6% Be doping shrank the tetragonal FeSe lattice equivalent to a physical pressure of 0.75 GPa. Using this flux-transport method of sample preparation, 6% of Be was the maximum amount of dopant achievable. At this maximal composition of $FeSe_{0.94}Be_{0.06}$, the lattice unit cell shrinks by 2.4%, $T_c$ - measured in the bulk via specific heat - increases by almost 10%, the $T_c$ vs pressure behavior shifts its peak $T_c^{onset}$ downwards by ~1 GPa, the high temperature structural transition around $T_S$ = 89 K *increases* by 1.9 K (in contrast to other dopants in FeSe which uniformly *depress* $T_S$), and the low temperature specific heat γ increases by 10 % compared to pure FeSe. Also, upon doping by 6% Be the residual resistivity ratio, $\rho(300K)/\rho(T\rightarrow 0)$, increases by almost a factor of four, while $\rho(300K)/\rho(T=T_c^+)$ increases by 50%.

**PACS index categories:** 74.70.Xa, 74.25.Bt, 74.25.F-

1. Introduction

FeSe has one of the lower superconducting transition temperatures ($T_c \sim 8$ K) of the iron based superconductors[1] (IBS) in its undoped, bulk form.  However, superconductivity
in FeSe is highly tunable.  With applied pressure (7 GPa) $T_c$ can approach[2] 37 K and, prepared as a monolayer on $SrTiO_3$, the $T_c$ of FeSe has reached, according to one report, 109 K,[3] the current record for high temperature superconductivity in IBS.

      S and Te are isoelectronic to Se, and are known to substitute on the Se site in FeSe. Doping FeSe with S ($FeSe_{1-x}S_x$), which shrinks the lattice size (i. e. chemical pressure), results[4] in a shallow peak in $T_c$ vs x at 10.7 K for x=0.11, while doping FeSe with Te ($FeSe_{1-x}Te_x$) expands the lattice and $T_c$ reaches[5] $\sim$14 K at $FeSe_{0.57}Te_{0.43}$.  Upon doping FeSe with either S or Te, the structural transformation, $T_S$, decreases monotonically in temperature at the rate of approximately 1.2 to 1.5 K per percent of dopant for both Te[5] and S[6].

      Various substitutions [7-8] for Fe in $Fe_{1-x}M_xSe$ have been tried, with degradation/destruction of $T_c$ with M=Ti, V, Co, Ni, Cu and Cr. The substitution of a nominal 2% of Mn in $FeSe_{0.5}Te_{0.5}$ (i. e. $Fe_{0.98}Mn_{0.02}Se_{0.5}Te_{0.5}$) has been reported [9] to increase $T_c^{onset}$ in zero-field-cooled magnetic susceptibility from 13.9 to 14.4 K.

      In the present work, single crystals of nominal composition $Fe_{1.063}Se_{0.9}Be_{0.1}$ were prepared using the flux growth method of Chareev et al. [10], where the excess Fe in the nominal concentration was to insure [11] the formation of the desired room temperature tetragonal structure phase.  At least with this preparation method, it was not possible to achieve further addition of Be into the lattice, and – as will be discussed below – the actual Be concentration in the nominal $Fe_{1.063}Se_{0.9}Be_{0.1}$ sample was found to be 6%.  After initial measurements on these crystals, as well as on crystals of the parent compound FeSe prepared in the same manner [10], crystals of composition $FeSe_{0.97}Be_{0.03}$ were grown in order to check for the systematic evolution of the properties with Be content.   The crystals with 6% Be as well as the parent compound were characterized using x-ray diffraction, energy dispersive spectroscopy (EDS), resistivity as a function of temperature, pressure and field, magnetic susceptibility, and specific heat.  For the $Fe_{1.063}Se_{0.97}Be_{0.03}$ crystals, resistivity and magnetic susceptibility were measured.  The covalent radius of Be (0.96 Å) is smaller than that of S (1.05 Å) and Se (1.2 Å) (radii for Te and Fe are 1.38 Å and 1.32 Å respectively), in agreement with the x-ray diffraction determination of the lattice parameters.  The reported rapid increase [2] in $T_c$ of FeSe with applied pressure ($T_c \sim 37$ K at 7 GPa) is linked to the extreme softness of the material, which has [2] a bulk modulus of only 31 GPa. The study of another smaller substituent besides S should provide insights as to the interesting properties of FeSe as well as the properties of IBS in general.

2. Experimental:

Chareev et al.[10] grew $Fe_{1.06}Se$ crystals in evacuated quartz tubes using a $KCl-AlCl_3$ flux, with the warm end of the tube (containing the Fe and Se starting material) at $\sim$430 ºC and the cold end of the tube (where the crystals formed) at $\sim$375 ºC.  This method was also followed in the current work.  Fe powder (99.998 % pure) and Se powder (99.999% pure), together with Be powder (MBE grade, 99.999% pure) in the proportion 1.063:0.9:0.1 (where the excess Fe concentration is to ensure [11] the tetragonal structure), were taken into a purified atmosphere glovebox, combined with the KCl and the hygroscopic $AlCl_3$, and placed in a 10 mm ID quartz tube. Without exposure to air, this tube was then sealed under vacuum.  In order to gain experience with the method (in particular the optimal ratio of flux to reactant), as well as to provide a good baseline for understanding the changes with Be-doping, pure $Fe_{1.063}Se$ and $Fe_{1.063}Se_{0.97}Be_{0.03}$ crystals (nominal concentrations) were also prepared using the same technique. In order to verify reproducibility, multiple furnace runs on all three compositions were carried out with the result that the crystals from differing tubes with the same nominal concentrations gave, within error bars, the same results.

The chemical composition of both the parent compound and the nominal $Fe_{1.063}Se_{0.9}Be_{0.1}$ was measured with a Hitachi S3400 scanning electron microscope operating at 20 kV. Energy dispersive x-ray spectroscopy (EDS) was checked for each sample on three areas (each ~ 80μm) and averaged. The results indicated a 1:1 stoichiometry in the parent compound (i. e. the excess Fe supplied in the growth tube does not enter into the final product crystals), and approximately $FeSe_{0.94}Be_{0.06}$ in the nominal composition $Fe_{1.063}Se_{0.9}Be_{0.1}$ sample. On the basis of this result, crystals with an intermediate nominal Be concentration, $Fe_{1.063}Se_{0.97}Be_{0.03}$, were prepared and characterized and indeed gave resistivity and magnetic susceptibility results (as discussed below) between those of the parent compound and $FeSe_{0.94}Be_{0.06}$

Our assumption in making these samples was that Be would go on the Se site, based on Be being closer in ionic radius to Se than Fe, as well as the existence of, e. g., the compound $FeBe_2$, i. e. Be is not always 2+ valent in compound formation. In order to check this assumption, single crystal X-ray diffraction was performed on crystals of FeSe and $FeSe_{0.94}Be_{0.06}$ and for both crystals we refined the occupancy of the Fe ($2b$ site) and Se ($2c$ site). The parent compound FeSe refines as 1:1, the same as the EDS result. For the crystal with 6% Be as determined by EDS, the Se site is deficient and in fact the sample composition refines as $FeSe_{0.94}$. Thus, we continued with our assumption that Be replaces Se, i. e. the Be-doped sample has the composition $FeSe_{0.94}Be_{0.06}$. (As will be seen below when we consider the extrapolated residual resistivity, $\rho(T\rightarrow 0)$, proportional to lattice defect scattering, this assignment is further substantiated by the lowered $\rho(T\rightarrow 0)$ and therefore lowered defect scattering in the Be doped sample, which would not obtain if the Se sublattice had 6% vacancies present.)

The single crystal x-ray diffraction gave a- and c-lattice parameters for the crystals of the parent FeSe compound equal to 3.7731 and 5.533 Å respectively, consistent with literature values. For the 6% Be-doped sample, the measured values of a and c were 3.7456 and 5.480 Å respectively. This indicated a shrinkage in the size of the lattice, as expected from the smaller Be radius, and guaranteed that the smaller Be did in fact replace either Fe or Se in the FeSe lattice and was not segregated as some second phase, leaving the lattice parameter unchanged, or enter the lattice interstitially, which would expand the lattice. Using the known bulk modulus of 31 GPa, this shrinkage corresponds to a physical pressure of ~0.75 GPa.

The high-pressure resistivity measurements were carried out in an OmniDAC gas membrane-driven diamond anvil cell from Almax-EasyLab. The cell was placed inside a custom, continuous flow cryostat built by Oxford Instruments. Optical access to the cell was provided through windows at the bottom of the cryostat and an optical fiber entering through a feed-through at the top, allowing pressure to be measured at low temperature. The pressure was calibrated using the fluorescence of the $R_1$ peak of small ruby spheres placed next to the sample[12]. One of the diamonds used was a designer diamond anvil[13] containing eight symmetrically arranged, deposited tungsten microprobes encapsulated in high-quality homoepitaxial diamond. This diamond had a tip diameter of ~180 $\mu$m, and the opposing anvil had a culet diameter of ~500 $\mu$m. Resistance was measured in the crystalline $ab$-plane using the four-probe Van der Pauw method with currents of ≤1 mA. Gaskets were preindented from 250 $\mu$m to ~30 $\mu$m thickness and were made of 316 stainless steel. Quasihydrostatic soft, solid steatite was used as the pressure-transmitting medium. The samples were cleaved from single crystals and had dimensions of ~70 $\mu$m x 70 $\mu$m x 10 $\mu$m.

The magnetic susceptibility was measured in a SQUID magnetometer from Quantum Design, while the resistivity and specific heat were measured in cryostats designed and built in the Department of Physics, University of Florida. The low temperature specific heat measurements were carried out using established techniques [14].

## 3. Results and Discussion

Resistivity and Magnetic Susceptibility:

The resistivities of the as-prepared single crystals of FeSe and $FeSe_{0.94}Be_{0.06}$ are shown in Fig. 1, with increases in $T_c^{midpoint}$ ($\approx 0.76$ K) and $T_c$ ($\rho \rightarrow 0$) (0.81 K) in the Be-doped over the parent compound as shown. (Data for $FeSe_{0.97}Be_{0.03}$, not shown for clarity, give an increase of $T_c$ ($\rho \rightarrow 0$) of 0.54 K.) The superconducting onset transition temperature of the two Be-doped ($FeSe_{0.97}Be_{0.03}$ and $FeSe_{0.94}Be_{0.06}$) samples as measured by magnetic susceptibility, more of a bulk measurement, (shown in Fig. 3) increase by ~0.3 and ~0.6 K vs the undoped parent compound respectively. (Specific heat of the parent compound compared to $FeSe_{0.94}Be_{0.06}$ is discussed below in the next section.)

The samples measured for magnetic susceptibility had masses in the range of 4.4 mg for the $FeSe_{0.97}Be_{0.03}$ up to 10 mg for the other two compositions. Thus, 50 G was chosen as the measurement field in order to achieve good signal to noise ratio in our MPMS system from Quantum Design. Both the FeSe and the $FeSe_{0.94}Be_{0.06}$ samples showed no signs (neither a large finite susceptibility above $T_c$ nor a saturation in the magnetization, M, measured vs field) of an Fe second phase. The $FeSe_{0.97}Be_{0.03}$ sample however (see Fig. 3) showed a slight elevation in $\chi$ above $T_c$, as well as a small moment (0.02 $\mu_B$/mole) that saturates at 0.3 to 0.4 T (M vs H data not shown.) Another collage of $FeSe_{0.97}Be_{0.03}$ crystals showed ~0.1 $\mu_B$/mole and a much more visible finite $\chi$ for temperatures above $T_c$. Thus, despite their good RRR value and the increase in $T_c$ vs the parent compound, the 3% Be-doped samples did, in some crystals, show a magnetic impurity phase. This may account for the larger resistivity at finite temperatures in the $FeSe_{0.97}Be_{0.03}$ (see Figs. 2 and 4).

The magnitudes of the reported values for $\rho$ depend on the geometry ($\rho=R*A/L$), where R is the measured resistance in ohms, A is the cross sectional area, and L is the distance between the contacts. Errors in this geometrical factor of 10% are inherent in the procedure.

It is interesting to note the large increase in the residual resistivity ratio, RRR, ($\equiv \rho(300K)/\rho(T \rightarrow 0)$) (see Fig. 1) from RRR=27 for FeSe to RRR=52 in the crystals doped with 3 %Be and RRR=102 for the crystals doped with 6% Be. (A second batch of $FeSe_{0.94}Be_{0.06}$ had RRR=105 and a second batch of $FeSe_{0.97}Be_{0.03}$ had RRR=84). If instead of $\rho(300K)/\rho(T \rightarrow 0)$, $\rho(300K)/\rho(T=T_c^+)$ is used, this ratio increases from 16.0 for pure FeSe to 24.7 for crystals doped with 6% Be. Doping FeSe with either S or Te on the Se site results in RRR values that either stay approximately the same [15-16] (S, x=0.1 and 0.18) relative to the parent FeSe compound, or *decrease* [16] (Te, x=0.25, 0.5). Similarly, doping FeSe on the Fe site also causes a *decrease* in RRR (Co [16], x=0.05, 0.1, 0.2; Mn [17], x=0.02).

Fig. 1: (color online) Resistivity of single crystals of FeSe and $FeSe_{0.94}Be_{0.06}$ vs temperature showing the shift of $T_c(\rho \rightarrow 0)$, $T_c^{mid}$ and $T_c^{onset}$ caused by the addition of Be. The midpoint, $\rho=0$

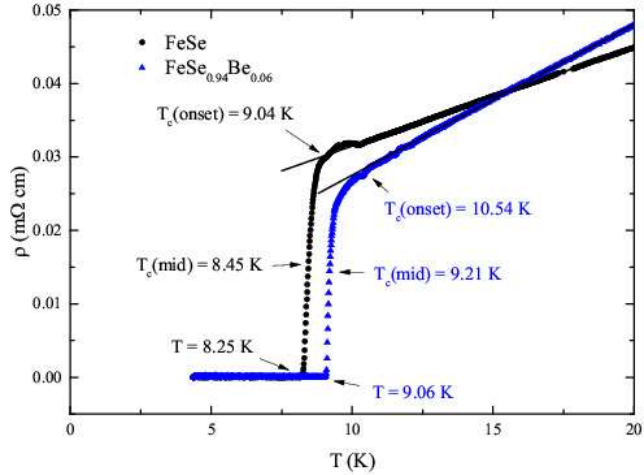

$T_c$, and RRR values of FeSe as determined by resistivity measurements of our single crystals are 8.45 and 8.25 K, and RRR=27 respectively, while the Be-doped sample has $T_c^{midpoint}$, $T_c(\rho \rightarrow 0)$ and RRR values of 9.21 K, 9.06 K, and 102 respectively. (For a single crystal of $FeSe_{0.97}Be_{0.03}$, the resistivity data – not shown – give $T_c(\rho \rightarrow 0)$=8.79 K, i. e. between the values for the parent compound and the 6% Be doped compound. The extrapolated normal state residual resistivity, $\rho(T \rightarrow 0)$, of the $FeSe_{0.97}Be_{0.03}$ is also between that of FeSe and the 6% Be doped compound.) The values for FeSe are comparable to literature values [18] for good quality single crystal samples, indicating the high quality of the undoped FeSe samples in the present work. However, it is not hard to find in the literature rather broadened resistive transitions where $T_c^{onset}$ for pure FeSe is significantly higher than our value of 9.04 K, see e. g. the ~13.5 K value in polycrystalline material of ref. [19] (where $T_c(\rho=0)$ was only 7.5 K). Note the apparent $\rho$ linear-with-T behavior for T>$T_c$ up to 20 K for both samples, and the larger linear-with-T slope for the Be-doped sample – see Fig. 2 for the extent of this linear behavior.

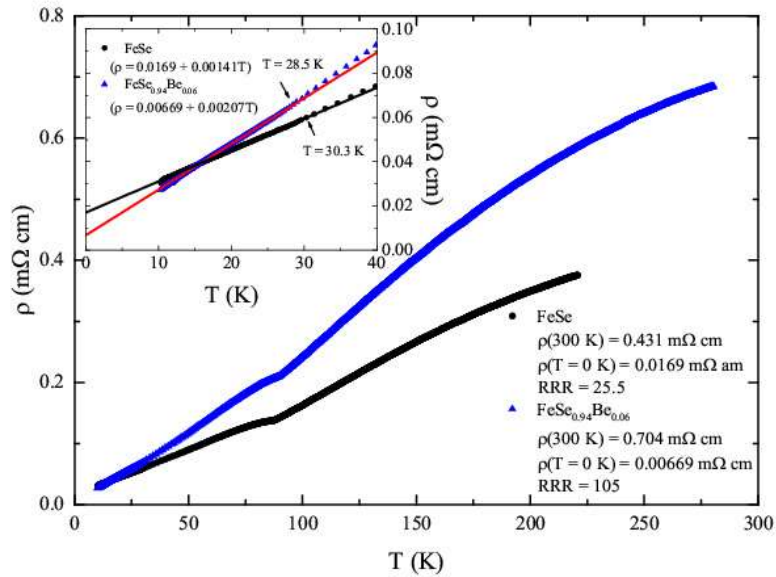

Fig. 2: (color online) The linear-with-T $\rho$ shown up to 20 K in Fig. 1 extends up to 28.5 K for the 6% Be-doped sample, and 30.3 K (i. e. essentially the same) for the parent FeSe compound as shown here in the inset. For the 3% Be-doped sample (not shown), the linear-with-T behavior extends up to 28.3 K, and the slope is 20% above that of FeSe. The crystals used for this figure were from the same batches as those for Fig. 1. In order to match the magnitude of $\rho$ for FeSe in Fig. 2 with that in Fig. 1, the estimated thickness of the Fig. 2 crystal was adjusted from an initial estimate of 0.007" to 0.0076". In the main part of the figure, the resistivities for FeSe and $FeSe_{0.94}Be_{0.06}$ are shown up to higher temperatures, with values for the $FeSe_{0.97}Be_{0.03}$ ~6% higher at room temperature. Although the extrapolated normal state resistivities down to T=0 for the Be doped samples, 3% (not shown) and 6%, are *lower* than that for the parent FeSe

compound – indicating less scattering from lattice defects – the behavior at high temperatures shows increased temperature dependent scattering (possibly from magnetic defects.)

It is worth stressing that the RRR (=$\rho$(300 K)/$\rho$(T$\rightarrow$0)) value for the FeSe$_{0.94}$Be$_{0.06}$ sample being almost four times that of the FeSe <u>primarily comes</u> from the *lower* residual resistivity extrapolated to T=0, $\rho$(T$\rightarrow$0), in the Be-doped vs that in the parent compound (0.0067 m$\Omega$-cm vs 0.017 m$\Omega$-cm, a factor of 2.5) rather than the higher $\rho$(300 K) (0.70 m$\Omega$-cm in Fe$_{1.063}$Se$_{0.94}$Be$_{0.06}$ vs 0.43 m$\Omega$-cm, a factor of 1.6, for the parent compound.) One possibility to explain the lower residual resistivity in FeSe$_{0.94}$Be$_{0.06}$ would be from the addition of carriers to the Fermi sea. However, as discussed in the next section, the specific heat results show only a 10% increase in the electronic specific heat coefficient $\gamma$, proportional to the dressed density of electronic states at the Fermi energy. Another possible source for the lower $\rho$(T$\rightarrow$0) values in the Be-doped compounds could be that the addition of Be caused improved sublattice order. Improved lattice order is in fact consistent with the high angle x-ray reflections from a collage of crystals oriented aligned with the x-ray beam in the c-axis direction, e. g. the [005] line in the 6% Be-doped sample is 20% narrower than seen for pure FeSe.

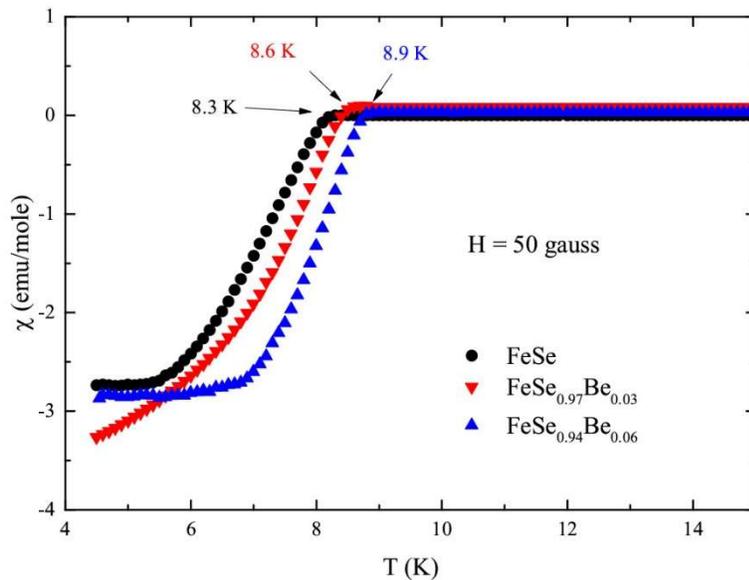

Fig. 3: (color online) dc magnetic susceptibility vs temperature (cooled in zero field, i. e. these are diamagnetic shielding, not flux expulsion, measurements) for FeSe, FeSe$_{0.97}$Be$_{0.03}$ and FeSe$_{0.94}$Be$_{0.06}$ single crystals. The increase in T$_c$ upon Be-doping is approximately the same as seen in the resistance data, although the transitions are shifted downwards in temperature as is typical when comparing resistive (corresponding to a one-dimensional pathway) and magnetic susceptibility (corresponding to two-dimensional shielding) results.

In any case, the lower residual resistivity $\rho$(T$\rightarrow$0) measured for the Be-doped crystals argues strongly (as discussed above) that there are not 6% vacant sites on the Se sublattice as would be one possible conclusion from the refinement of the x-ray diffraction results, but that in fact the doped Be resides on the Se sublattice.

At present there is no explanation for why Be might improve sublattice order, leaving this as an interesting puzzle for further investigation.

The resistivities of FeSe, FeSe$_{0.97}$Be$_{0.03}$ and FeSe$_{0.94}$Be$_{0.06}$ in the vicinity of where the tetragonal – orthorhombic structural transition takes place at T$_S$ ~ 89 K are shown in Fig. 4. The structural transition shifts upwards in temperature approximately 1.9 K with the addition of 6 % Be. This increase in T$_S$ tracking an increase in T$_c$ in Be-doped FeSe is contrary [1] to the great

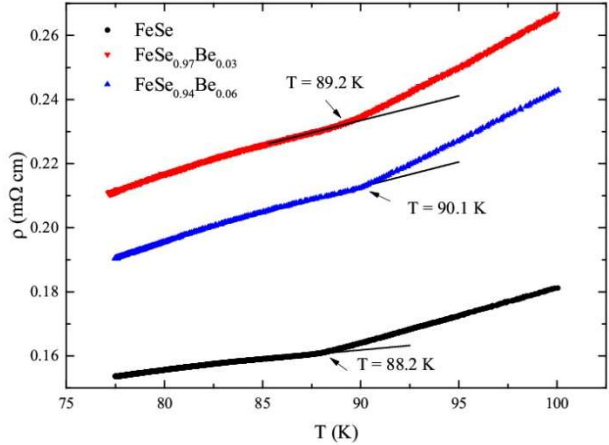

Fig. 4 (color online): Resistivity of Fe$_{1.063}$Se, FeSe$_{0.97}$Be$_{0.03}$ and Fe$_{1.063}$Se$_{0.94}$Be$_{0.06}$ single crystals at the tetragonal to orthorhombic phase transition at T$_S$. As seen, the scattering at finite temperatures in the Be doped samples (perhaps from magnetic impurities – see Fig. 3) causes a higher resistivity than in the parent compound at the structural ordering transition.

majority of results in IBS, where an increase of doping drives T$_c$ upwards on a superconducting 'dome', while at the same time the structural transition temperature T$_S$ is *depressed*. Also, as discussed above in the Introduction, doping FeSe with either S [6] or Te [5] reduces T$_S$. If we consider the addition of Be simply as equivalent to chemical pressure, based on the shrinkage of the c- and a-axis lattice parameters, this increase in T$_S$ is contrary to measurements of T$_S$ under pressure where T$_S$ *decreases* quite rapidly [20] with pressure, at about 25 K/GPa.

One possible way to understand this behavior is to note the fact that the RRR is significantly increased in the Be-doped samples, i. e. consistent with an improvement in the sublattice order. This would improve local chemical and electronic uniformity [21], resulting in a higher structural ordering temperature. Such an improvement of T$_S$ with increasing RRR was seen in ref. [18]'s study of vapor grown FeSe crystals. As we will see below when comparing the specific heat of Be-doped FeSe to that of S-doped FeSe, the addition of Be is not simply equivalent to chemical pressure in the specific heat either.

One of the important characterizations of FeSe$_{0.94}$Be$_{0.06}$, with its lattice already 'pre-compressed' via the addition of the smaller Be, is how its superconducting transition temperature responds to pressure. T$_c^{onset}$ as a function of pressure is shown in Fig. 5 for *both* FeSe$_{0.94}$Be$_{0.06}$ and for the undoped FeSe prepared using the same [10] method in this work. Fig. 6 shows a sample determination of how T$_c^{onset}$ is determined at 5.3 GPa.

Fig. 5 shows that the increase in T$_c$ in the Be-doped FeSe does not lead to a higher eventual peak T$_c$ achieved with pressure, but rather just shifts the T$_c$ vs P curve to slightly lower pressure, i. e. the peak T$_c^{onset}$ is reached approximately 1 GPa earlier. Thus, numerically the addition of 6% Be to FeSe is approximately equivalent to a 1 GPa pressure shift in the T$_c$ vs P curve. The 'chemical pressure' of Be substitution, determined from the shrinkage in the lattice parameter together with the known bulk modulus, as discussed above is approximately equivalent to 0.75 GPa, roughly equivalent to this high pressure result.

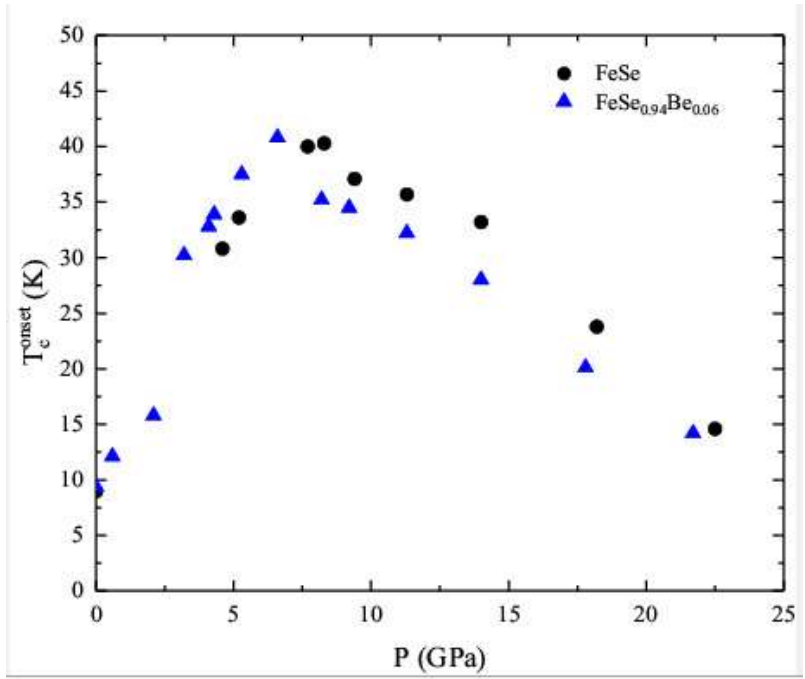

Fig. 5: (color online) $T_c^{onset}$ vs pressure in a diamond anvil cell for $FeSe_{0.94}Be_{0.06}$ and FeSe. $T_c^{onset}$ exceeds 40 K at about 7 GPa for $FeSe_{0.94}Be_{0.06}$, while this peak $T_c$ is shifted higher by ~ 1 GPa for the undoped $Fe_{1.063}Se$. The data shown here for pure FeSe are similar to those taken by Sun et al.[22], except peak $T_c^{onset}$ in their data is ~2-3 K lower, and their data extended only to 10 GPa. Suppression of $T_c$ with field applied in the c-axis direction up to 9 T was also studied (not shown) at ambient pressure. $H_{c2}(0)$ shifts upward by 1 T in the 6% Be-doped compound vs the undoped parent compound.

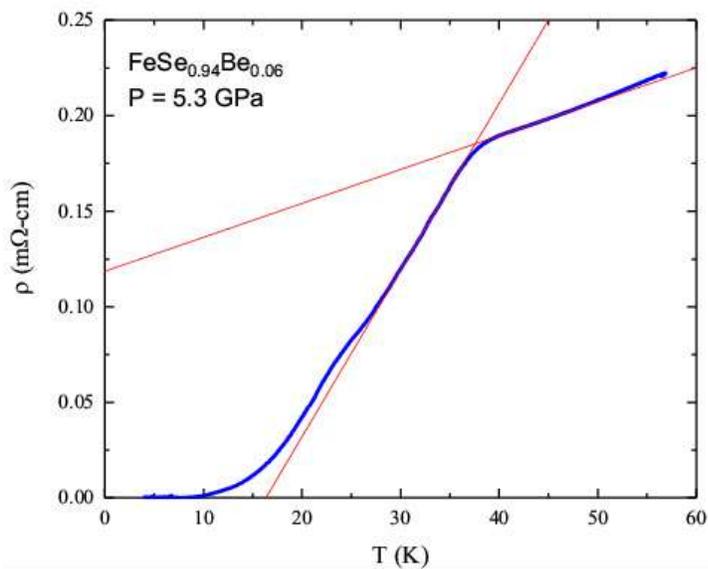

Fig. 6: (color online) Although the resistive transition broadens significantly in the diamond anvil pressure cell (compare the 10-90% width at 5.3 GPa of ~15 K shown here with ~1 K at zero pressure, see Fig. 1), the onset is easy to accurately determine, as shown here. This width is similar to that observed in pure FeSe at 5 GPa in our own work (not shown), in pure FeSe at 4.15 GPa by Masaki et al.[23], who used an indenter cell, but is signicantly broader than the data for FeSe of Sun et al. [22], who used a much more hydrostatic clamp type cubic anvil cell. Since $T_c$ exhibits a strong pressure dependence, small pressure gradients can produce a significant spread in $T_c$ values, resulting in a broad transition. Hence, it is not surprising that the more hydrostatic measurements of Sun et al. exhibit a narrower transition width. Note that the resistivity at 40 K in 5.3 GPa shown here in the Be doped FeSe is about a factor of two higher than in zero pressure (Fig. 2). This is an interesting contrast to the slight decrease or constant behavior seen in the same pressure regime in the parent compound FeSe in the data of refs. 22 and 23.

Specific heat:

The specific heat of single crystals of $FeSe_{0.94}Be_{0.06}$ and FeSe is shown in Fig. 7. We wish to stress that the comparison between the specific heat of these two samples is fairly precise: the relative precision should be better than ±3 % since the same mass sample with the same addenda on the same platform in the same apparatus was measured. Plus, we have intentionally – rather than taking literature data[4,24-25] for the specific heat of FeSe (where reported γ values vary by at least 10% even when restricting the samples to only single crystals grown by the same Chareev et al.[10] flux growth technique) – grown FeSe single crystals with the same method as for our $FeSe_{0.94}Be_{0.06}$ crystals to provide the best intercomparability. Since the bulk $T_c^{onset}$ determined by specific heat of our $FeSe_{0.94}Be_{0.06}$ single crystals (as well as $T_c^{midpoint}$ determined by resistivity, RRR, and $T_c^{onset}$ determined from dc susceptibility) are higher than the literature values [4,24-25], as well as our values, for undoped FeSe, there is no doubt that the doped Be has an effect.

As can be seen in the specific heat data (fit to $C/T = \gamma + \beta T^2$) of Fig. 7, doping Be into FeSe – in addition to raising the bulk $T_c^{onset}$ by ~0.8 K vs our comparison crystals of FeSe - also increases the specific heat γ value by about 10%. The slight (~3%) change in β ($\propto 1/\Theta_D^3$) gives a minor change in the respective Debye temperatures, $\Theta_D$, (215 K for the Be-doped and 218 K for the pure FeSe) which can be seen in the slightly faster rise in C/T for $T > T_c$ in Fig. 7. However, this slight change in the Debye temperature is within the respective error bars of the two sets of data. Literature values [4,24-25] for the Debye temperature of FeSe single crystals are ~ 210 K. On the other hand, the increase in the specific heat γ in the Be-doped compound, although only 10% in size, relative to the undoped FeSe is – with the effort spent in this work to achieve internal precision – significant.

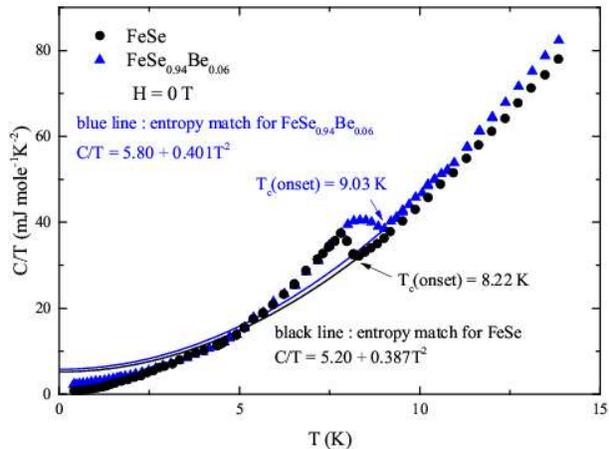

Fig. 7: (color online) The low temperature specific heat divided by temperature, C/T, vs temperature for a collage of ~10 flat platelet single crystals of $FeSe_{0.94}Be_{0.06}$, mass ~ 10.5 mg, and of a collage of ~10 flat platelet single crystals of FeSe, mass ~ 10.5 mg. (A second collage of $FeSe_{0.94}Be_{0.06}$, using crystals from the same growth tube, was also measured (not shown) and gave similar results, with the exception that the transition was broader and started at 9.27 K.) Matching the measured superconducting state entropy for each sample up to $T_c$ with the respective extrapolated normal state entropy up to $T_c$ for each sample gives the $C/T = \gamma + \beta T^2$ dependences shown.

4. **Summary and Conclusions**

Introducing Be into FeSe produces a slight increase in $T_c$ – comparable to that seen [4] with the introduction of a few percent of S. While $T_c$ shifts upwards by 10% in $FeSe_{0.94}Be_{0.06}$, the structural transition around 89 K shifts upwards by 1.9 K (1 K for 3% Be) and the $T_c$ vs pressure curve shifts to lower pressures by ~1 GPa. X-ray determination of the room temperature tetragonal lattice parameters, coupled with the known bulk modulus of 31 GPa for FeSe, indicates a shrinkage of the lattice corresponding to ~0.75 GPa. The residual resistivity ratio, possibly indicative of the perfection of sublattice ordering, increases by ~ factor of four (two) with the addition of 6% (3%) Be. This large increase in RRR is surprising since doping with a foreign atom in a lattice typically *enhances* scattering. The low temperature specific heat indicates, within a rather small error bar, that the introduction of 6% Be does not significantly change the lattice stiffness (Debye temperature) in FeSe, while the electronic specific heat coefficient increases from 5.2 to 5.8 mJ/molK$^2$. This contrasts with literature results [4] for FeSe$_{0.91}$S$_{0.09}$, $T_c$=10.1 K, where the γ value reported was a *decrease* to 4.9 mJ/molK$^2$, vs the value reported [4] in the same work for pure FeSe of γ=5.3 mJ/molK$^2$. Thus, it would be reasonable to conclude that the differences between S and Be doping (both smaller than Se) relate to the different electronic properties of the dopants. Work is underway to introduce Be into both FeSe$_{1-x}$S$_x$ and FeSe$_{1-x}$Te$_x$ to further investigate the utility of Be doping for understanding the superconductivity in FeSe and its derivatives.

Acknowledgements: Work at Florida by J. S. K. and G. R. S. performed under the auspices of Basic Energy Sciences, U. S. Department of Energy, contract. No. DE-FG02-8645268. Development of *in situ* pressure tuning equipment partially supported by The National High Magnetic Field Laboratory User Collaboration Grants Program. The National High Magnetic Field Laboratory is supported by National Science Foundation Cooperative Agreement No. DMR-1157490 and the State of Florida. Measurements supported by DMR-1453752, designer diamond anvils supported by DOE-NNSA Grant No. DE-NA0002928 and under the auspices of the U.S. Department of Energy by Lawrence Livermore National Laboratory under Contract DE-AC52-07NA27344. Chemical analysis and diffraction work at ORNL was supported by the U.S. Department of Energy, and the Scientific User Facilities Division, Office of BES, and the Office of Science, Basic Energy Sciences, Materials Sciences and Engineering Division. Enlightening conversations with Peter Hirschfeld and Sergei Budko are gratefully acknowledged.